# Integration of the Packetdrill Testing Tool in INET

Irene Rüngeler
Münster University of Applied Sciences
Dept. of Electrical Engineering and Computer Science
Bismarckstrasse 11
D-48565 Steinfurt, Germany
Email: i.ruengeler@fh-muenster.de

Michael Tüxen
Münster University of Applied Sciences
Dept. of Electrical Engineering and Computer Science
Bismarckstrasse 11
D-48565 Steinfurt, Germany
Email: tuexen@fh-muenster.de

*Abstract*—Google released in 2013 a script-based tool called `packetdrill`, which allows to test transport protocols like UDP and TCP on Linux and BSD-based operating systems. The scripts defining a test-case allow to inject packets to the implementation under test, perform operations at the API controlling the transport protocol and verify the sending of packets, all at specified times. This paper describes a port of `packetdrill` to the INET framework for the OMNeT++ simulation environment providing a simple and powerful method of testing the transport protocols implemented in INET.

## I. INTRODUCTION

The complexity of software involves a great vulnerability to bugs. As a consequence, testing becomes more and more vital, but also time consuming. The possibility to automate tests is therefore a must.

The INET framework includes a test suite [1], that features regression, module and validation tests. The user can define tests that are validated via fingerprints or the comparison of parts of the output with a regular expression. These tests can run individually or in a batch job. The test suite is very well suited to test new features for a protocol that expect a certain observable output, but it is hard to test scenarios where you have specific interactions with the applications or specific patterns in packet reception.

As our focus lies on transport protocols we needed a testing tool with special emphasis on the transport layer and its interaction with the application via the socket Application Programming Interface (API). Our preferred transport protocol is Stream Control Transmission Protocol (SCTP), but a more versatile test tool that can be used to also test other transport protocols like User Datagram Protocol (UDP) and Transmission Control Protocol (TCP) would be of broader interest.

Cardwell [2] published the open-source scripting tool `packetdrill`, that "enables testing the correctness and performance of entire TCP/UDP/IP network stack implementations, from the system call layer to the hardware network interface, for both IPv4 and IPv6" [3]. `packetdrill`, developed at Google, is what we were looking for. It combines the following features:

- It is possible to call system calls and compare the expected packets with the real packets created by the network stack.
- The timing is supervised and wrong timing is a reason to fail a test.
- Packets with unsuitable parameters can be injected in the stack and the reaction of the implementation observed.
- A suite of test scripts can be designed and used for regression.

But, it is only suitable to test kernel stacks, since it explicitly uses the socket API and deals with real interfaces.

Our intention was therefore to port `packetdrill` to the INET framework in order to benefit from the same testing features as the kernel implementations.

Yet, another advantage would be to compare the simulation with the Linux and BSD kernel implementations and even use the same test scripts.

To achieve these goals, support for TUN interfaces was added to the INET framework. These interfaces are used to handle the packets for testing. Furthermore, the `packetdrill` application has to be ported from a procedural to an object oriented program. Finally, SCTP support has to be added to support all transport protocols currently being implemented in the INET framework. We are in the process of contributing this work to the public INET repository and hope that our code will be accepted soon.

This paper describes our approach and is structured as follows: In Section II the features of `packetdrill` as implemented by Google will be introduced. The porting of this tool, its architecture and necessary new components in INET, are explained in Section III. Section IV concludes the paper and gives an overview of the future work.

## II. THE GOOGLE PACKETDRILL TESTING TOOL

The Google `packetdrill` testing tool was created to ease the process of testing during development, debugging and regression [2]. The scripts can be tailored to test exactly the packet flow that is needed being able to leave out parameters that are not of interest for the test. The script syntax is similar to that of *tcpdump* for packet descriptions and *strace* for system calls. *tcpdump* is a tool for capturing and analyzing packets which is available on most Unix systems. *strace* is a debugging tool showing issued system calls available on Linux systems.

Like every other test tool `packetdrill` has a System Under Test (SUT) and an application that handles the tests. For each test a script is written that features the expected message flow.

```
 0.000    socket(..., SOCK_STREAM, IPPROTO_TCP) = 3
 0.000    setsockopt(3, SOL_SOCKET, SO_REUSEADDR, [1], 4) = 0
 0.000    bind(3, ..., ...) = 0
 0.000    listen(3, 1) = 0
 0.100  < S   0:0(0) win 32792 <mss 1460, sackOK, nop, nop, nop, wscale 7>
 0.100  > S.  0:0(0) ack 1 <...>
 0.200  < .   1:1(0) ack 1 win 257
 0.200    accept(3, ..., ...) = 4
 0.300  < F.  1:1(0) ack 1 win 260
+0.000  > .   1:1(0) ack 2
 0.320    close(4) = 0
+0.000  > F.  1:1(0) ack 2
+0.000  < .   2:2(0)  ack 2 win 32792
```

Fig. 1. Example of a `packetdrill` script

The script in Figure 1 shows a TCP server as the SUT, which accepts a TCP connection. Then the client closes the connection. Initially a TCP socket is opened. Then a socket option is set, the socket is bound and `listen` is called. The server awaits a SYN from the client and answers with a SYN-ACK. The handshake is concluded by an ACK from the client. The characters '<' and '>' denote the direction of the packet. '<' indicates that the packet is injected, i. e. it is sent via the link layer towards the stack being tested. This way mimicks the client's reaction. It is possible to send packets with wrong values in order to test the SUT's behavior. To be able to inject a packet, a virtual network TUNnel (TUN) interface is used. It is automatically opened by `packetdrill` and has a direct connection to the application. Packets starting with '>' are created to be compared to the real packets being sent by the stack being tested. The real packets are created by the stack either as a reaction to a system call or a message from the client. The outgoing real traffic is read using the `libpcap` [4] and compared to the expected packets. For example, the outgoing SYN-ACK is created by `packetdrill` and stored to be compared to the sniffed SYN-ACK. In the course of the script the FIN bit is sent by the client and the server invokes the `close` call.

When `packetdrill` is started the script is parsed, the packets are created and stored either to be injected at a specified time or to be compared later. Then a second thread is started, in which the events (system calls and packets) are processed one by one according to their time stamps.

There are six ways to set the time a packet is expected to be processed. It can be absolute, relative (e.g. +0.1) or in a time range (e.g. 0.1 ∼ 0.2). Wildcards can be used or time boundaries for blocking system calls specified. The expected time of an outgoing packet is compared with the real time of a live packet. When the difference is greater than a configured time tolerance, the test has failed. In the case of injected packets the time stamp sets the injection time.

`packetdrill`, as described in [2] and [3] can be used in two different modes, the local and the remote mode. In the local mode only one host is needed, whereas in remote mode two hosts are needed. In local mode a logical TUN interface is used allowing the testing of the transport and network stack, but not capturing any effects (like offloading checksum computations, segmentation and reassembly, etc) a real network interface card and its driver produce. However, in both modes the same logical network setup is used which is depicted for IPv4 in the following Figure 2.

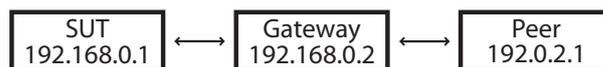

Fig. 2. IPv4 Address Configuration for `packetdrill`

The SUT is using a private IPv4 address and communicates with a peer using an address reserved for testing via a gateway in the same network as the SUT.

Besides system calls and packets Google `packetdrill` supports shell and python commands. To configure the SUT using *sysctl* or assess the state of the machine, a shell command can be included in the script. To print information or use asserts, python snippets are added to the script.

To select one of the supported network layer protocols IPv4 or IPv6 a command line flag can be set. Thus, there is no need to change the script.

In addition to the transport protocols UDP and TCP, we implemented support for SCTP.

### III. PORTING PACKETDRILL TO INET

As the transport layer, especially SCTP, is our main research subject `packetdrill` is the ideal testing tool for us. The first idea was to only add SCTP support to Google `packetdrill`. But as we always keep the SCTP implementation in INET up-to-date with kernel SCTP, we decided to port `packetdrill` to INET. Our goals were to use the same scripts for the kernel version and the simulation and to reuse as much code as possible. As Google released `packetdrill` under the GPL2 integration of it in INET is not a problem.

#### A. Addition of TUN Interfaces

The TUN interface, as shown in Figure 3, is integrated like any other interface to the NodeBase node. It is activated via a parameter in the omnetpp.ini configuration file. The specific feature of this interface is its direct connection to an application. Thus, traffic arriving from the network layer at the TUN interface is diverted to an application to be processed instead of to another host. To get a better overview of the packets traversing the interface, the pcap recorder can be used to trace the traffic. The TUN interface has already been accepted into the integration branch of the inet-framework.

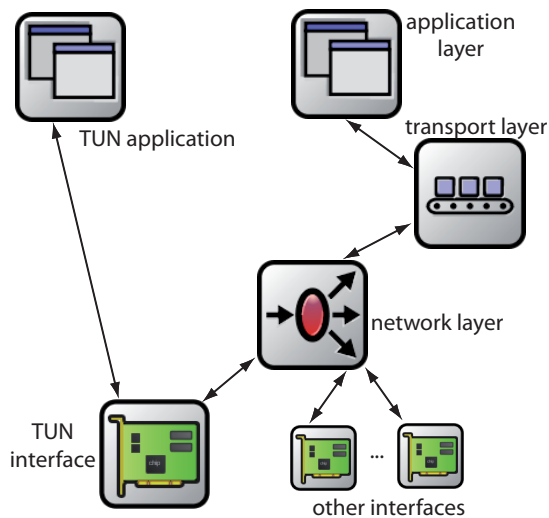

Fig. 3. TUN Interface and its Application

Besides this general behavior of a TUN interface, we had the specific requirement to supervise the traffic going up and down the protocol stack because we wanted to inject packets, i.e. going up, and compare the real traffic to the expected packets, i.e. going down. The solution was an application for both the TUN interface and the transport layer as shown in Figure 4. The new PacketdrillHost has gates to the TUN interface and to UDP, TCP and SCTP on the transport layer.

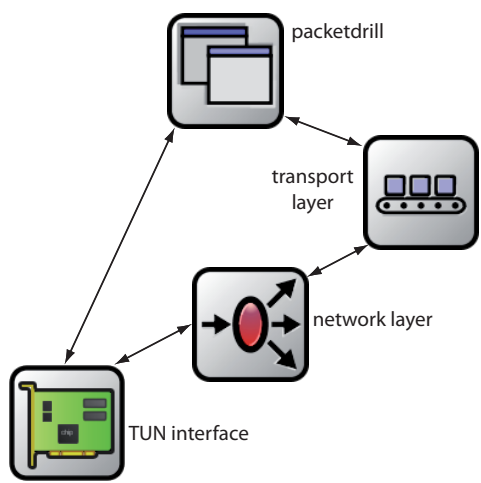

Fig. 4. `packetdrill` Host

### B. Software Adjustments and Enhancements

Central parts of Google `packetdrill` are the parser and the lexer that verify the script and generate the list of events to be processed in the course of the test. We did not want to reinvent the wheel. Therefore, we decided to keep as much of this mechanism as possible. But still we had to adjust the parser to the requirements of INET. As Google `packetdrill` is written in C while INET is implemented in C++, a lot of new classes had to be designed to substitute the structures and other data types used in Google `packetdrill`.

To meet the timing requirements of the tests, a strict sequence of the events had to be established. As we only have one thread working in INET, events cannot be processed in parallel. The relative time of an event can only be determined, after its predecessor has been processed. Event timers have to be scheduled to start the sending or receiving of packets at the correct time. The events are numbered not to allow the processing of an event before the predecessor is finished. When simulated packets are travelling down the stack they do not require simulation time. Therefore, it is possible to expect a packet at `packetdrill` via the TUN interface in zero seconds. But still, different arrival times are possible, if, for instance, retransmission times are expected. If the timing behavior is not in the focus of the test, wildcards can be specified to make clear that any time is accepted.

Outgoing packets have to be created and stored until the corresponding live packet arrives via the TUN interface. After `packetdrill` has verified that the live packet arrived in the expected time range, the live and the stored packet have to be compared starting at the IP layer. Injected packets have to be filled exactly, i.e. a complete IP datagram has to be defined in the script, because the packet is sent directly up the stack mimicking the arrival of a packet from a remote host. The definition for outgoing packets is not as strict. Parameters that are not of interest for the test, may be omitted. Yet, it must be possible to compare any parameter that could be used in the protocol. If a parameter is not equal in both packets, a termination exception is thrown and the test is finished.

As mentioned before, system calls trigger commands that are sent from the application layer to the transport layer where they initiate a predefined reaction. The communication between these layers is defined in the socket API of the respective transport protocol. A frequently used system call is *setsockopt()*, which sets options for the created socket (see script in Figure 1). Up to now it was not possible for TCP and SCTP in INET to set parameters via a system call. Options were mostly declared in the .ned files of the application and the modules of the transport protocol and set in the configuration file. But as the scripts for real implementations should be usable in the simulation, a mechanism had to be realized to hand socket option down from the application to the socket and up from the transport level to the socket in order to be able to overwrite predefined values. The actual values are then stored in the main module of the transport layer.

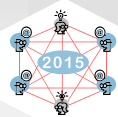

For UDP and TCP the grammar in Google `packetdrill` was taken, but for SCTP [5] a new grammar with new rules had to be written. SCTP's packets consist of a common header and different chunk types, each with a header and parameters of its own. All possible combinations had to be respected in the parser and the lexer and for all chunks functions had to be added to compare the content.

Once the main structure of a protocol is integrated it will not be difficult to add new protocol features, like new chunk types.

*C. Test Cases*

As the scripts are embedded in the INET test suite, their specification follows the same rules as other module tests. The `packetdrill` tests always consist of a script, a .ned file, a routing file and a configuration file like the following.

```
%description:
SUT is server. It accepts a connection.
The peer closes the connection.
%#---------------------------------------------
%inifile: omnetpp.ini

[General]
network = PacketDrillTcp
debug-on-errors = true
ned-path = .;../../../../../src;../../lib

**.scriptFile="../../lib/openPassive.pkt"
**.pdhost.numTcpTunApps = 1
**.hasTun = true

**.startTime = 2s

**.pdhost.routingFile = "../../lib/pdhost.mrt"
**.pdhost.numPcapRecorders=1
**.pdhost.pcapRecorder[0].pcapFile
    ="openPassive.pcap"
**.pdhost.pcapRecorder[0].moduleNamePatterns
    ="tun"
**.pdhost.pcapRecorder[0].sendingSignalNames
    ="packetSentToUpper"
**.pdhost.pcapRecorder[0].receivingSignalNames
    ="packetReceivedFromUpper"

**.pdapp.dataTransferMode = "bytecount"
**.tcp.mss = 1460
**.tcp.sackSupport = true
**.tcp.windowScalingSupport = true
**.tcp.windowScalingFactor = 6
**.tcp.advertisedWindow = 29200
**.tcp.useDataNotification = true

%#---------------------------------------------
%not-contains: test.out
Packetdrill error:
%#---------------------------------------------
```

The `%description` characterizes the test that is configured in the following `%inifile`. The script file is specified, the protocol, the routing file and the parameters for the pcapRecorder. After the protocol specific parameters the condition for a successful test is stated. The cause for the above mentioned termination exception always starts with "Packetdrill error:" and can thus be used as an error indication.

*D. Limitations of the simulation in relation to Google `packetdrill`*

Looking at Google `packetdrill` as a model for the INET version, not all features have been implemented. The following limitations apply currently:

- Remote mode unsupported
  The remote mode allows also for testing any interactions of the network interface card with the transport layer. Since the network interfaces in the INET framework do not perform any kind of offload or other interaction with the transport layer, the support of the remote mode was not a priority.
- Python snippets, shell code and command line arguments unsupported
  In Google `packetdrill` command line arguments are interpreted and python snippets can be included as well as shell code. In INET the command line arguments could be realized by adding configurable parameters to the .ned files and thus configure them in the test file. It would probably mean a major enhancement to INET to allow the interpretation of python snippets and shell code. As we have no use case for this feature, we did not implement it initially.
- Blocking system calls unsupported
  Since INET does not support blocking system calls there was no need to implement them.
- *getsockopt()* unsupported
  The system call *getsockopt()* as a counter part to *setsockopt()* can be used to query socket options. Initial test cases had no need for *getsockopt()* support. Therefore, is was postponed.
- Explicit address handling unsupported
  In Google `packetdrill` it is possible to set the addresses of the sender and receiver as part of the event header. This functionality has been postponed until use cases for address handling arise.
- IPv6 unsupported
  Up to now only IPv4 is supported since we are focusing on the transport layer, but IPv6 will be included in a future version, as it becomes more and more important.

IV. CONCLUSION AND OUTLOOK

So far we have ported the main features of Google `packetdrill` except for those mentioned in the last section. Yet, there are some features that are missing in both Google `packetdrill` and INET. The support for SCTP has already been implemented in both versions. One of SCTP's outstanding features is multihoming, the parallel use of more than one IP address over one connection. It is desirable to test the handover of traffic from one path to another and the simultaneous use of several paths. Therefore, it needs to be added to `packetdrill`. After completing `packetdrill`, future work will focus on the development of test suites for transport protocols being available in the INET framework. These scripts should allow testing the transport stacks in the INET framework and on Linux and BSD based operating systems.